\documentstyle[12pt]{article}
\input epsf.tex
\input FEYNMAN

\newcommand{\be}{\begin{equation}}
\newcommand{\ee}{\end{equation}}
\newcommand{\ba}{\begin{eqnarray}}
\newcommand{\ea}{\end{eqnarray}}
\newcommand{\pa}{\partial} 

\def\bea{\begin{eqnarray}}
\def\eea{\end{eqnarray}}

\def\bb#1{\hbox{\mybb#1}}

\def\R4{\bb{R}^4}
\font\mybb=msbm10 at 12pt
\def\tr{{\rm Tr}}

\def\tr{{\rm Tr}}

\def\D{{\cal D}}
\def\unita{{1 \kern-.30em 1}}

\begin{document}

\begin{titlepage}
\begin{flushright}
{IFUM 547/FT}\\
\end{flushright}
\vskip 1mm
\begin{center}
 
{\large \bf  Feynman rules and $\beta$-function for the BF Yang-Mills 
theory }\\ 
  
\vspace{1cm}

{\bf Maurizio Martellini}

\vskip 0.1cm
{\sl Dipartimento di Fisica,Universit\`a di Milano,}\\ 
{\sl I.N.F.N.\ - \ Sezione di Milano} \\ 
{\sl and}\\ 
{\sl Landau Network  at ``Centro Volta'', Como, ITALY}\\  
{\it martellini@vaxmi.mi.infn.it}\\ 
\vskip .2cm
 
{\bf Mauro Zeni}
 
\vspace{0.1cm}
{\sl Dipartimento di Fisica, Universit\`a di Milano \\
and \\  
I.N.F.N. \ - \  Sezione di Milano, \\
Via Celoria 16 \ \ 20133 \ Milano \ \ ITALY}\\ 
{\it zeni@vaxmi.mi.infn.it}

\end{center}
\abstract{
Yang-Mills theory in the first order formalism appears as the 
deformation of a topological field theory, the pure BF theory. We 
discuss this formulation at the quantum level, giving the Feynman rules 
of the BF-YM theory, the structure of the renormalization 
and checking  its
{\it uv}-behaviour in the computation of the 
$\beta$-function which agrees with the expected result.
}
\vfill
\end{titlepage}
\setcounter{section}{0}
\section{Introduction}
\setcounter{equation}{0}
\addtolength{\baselineskip}{0.3\baselineskip} 

Gauge theories, which play a central role in our understanding of 
high energy interactions, are usually described in terms of the Yang-Mills 
action. In this letter we consider the first order formulation of Yang-Mills 
theory, in which an auxiliary tensor field $B$ couples to the physical degrees 
of freedom of the gauge theory. This formulation, 
which has been used in \cite{fmz,mz} 
to introduce an explicit representation of the `t Hooft algebra \cite{thooft1}, 
makes closer the connection between Yang-Mills theory and topological field 
theories of BF type \cite{horouno,blau}; we will call this formulation 
BFYM theory. 
We give  Feynman rules for BFYM theory and discuss the structure of one 
loop divergent diagrams and renormalization and check the $uv$-behaviour of 
the theory computing the $\beta$-function which turns out to agree with the 
expected value; some of these results have been anticipated in \cite{mz}.
 
The first order form of pure euclidean Yang-Mills theory 
is described by the action functional
\ba 
S_{BFYM} &=& \int \tr [iB\wedge F +g^2B\wedge *B] \nonumber \\
&=& \int d^4x ({i\over 2}\varepsilon^{\mu\nu\alpha\beta}
B^a_{\mu\nu}F^a_{\alpha\beta}
+g^2B^a_{\mu\nu}B^{a\mu\nu})  \quad ,
\label{uno.1}
\ea 
where $F=F^a_{\mu\nu}dx^\mu\wedge dx^\nu \hat T^a$ is the usual field strength,
$D\equiv d +i[A,\cdot ]$ and $B$ is a Lie valued 2-form \cite{hr}. 
The generators of the $SU(N)$ Lie 
algebra in the fundamental representation are normalized as 
$\tr \hat T^a \hat T^b=1/2\delta_{ab}$ 
and $*$ is the Hodge product for a $p$-form. 
The field equations of (\ref{uno.1}) are 
\ba
F&=& 2ig^2*B\quad ,\nonumber \\
DB&=&0\quad .
\label{uno.2}
\ea
The standard YM action is recovered performing path integration over B 
or by using equations (\ref{uno.2}) in (\ref{uno.1}). 
Therefore the BFYM action (\ref{uno.1}) 
is on-shell equivalent to YM theory and its classical gauge invariance 
is given by
\ba 
\delta A &=& D\Lambda_0  \quad ,\nonumber \\ 
\delta B &=& i[\Lambda_0,B]  \quad .
\label{uno.3} 
\ea 
The question arises whether the two formulations are equivalent at the 
quantum level and to which extent this equivalence holds. Note that 
off-shell the field $B$ is not constrained by any Bianchi identity and 
this fact has been related to the presence of magnetic vortex lines 
in the vacuum of the theory  
in \cite{fmz}, in the picture of the dual superconductor 
vacuum \cite{mandel,thsc}. 

In the limit of vanishing coupling, $g\to 0$, the action (\ref{uno.1})  
flows in the pure BF theory \cite{horouno,blau} 
which is known to give a topological 
field theory, 
\be  
S_{BF}=i\int \tr [B\wedge F]\quad .
\label{uno.4} 
\ee 
Indeed the action (\ref{uno.4}) has a second gauge symmetry, namely 
\ba 
\tilde\delta A &=& 0 \quad ,\label{uno.5} \\ 
\tilde\delta B &=& D\Lambda_1 \quad ,
\nonumber 
\ea 
where  $\Lambda_1 $ is a 1-form. The presence of this 
``topological'' symmetry cancels out any local degree of freedom from the 
theory (\ref{uno.4}).  

Therefore YM theory in the BF formulation appears as a deformation of the
topological field theory (\ref{uno.4}); the $g^2B^2$ 
term which allows gaussian 
integration in (\ref{uno.1}) is  an explicit breaking term for the symmetry
(\ref{uno.5}).  Since pure BF theory is known to be a finite theory 
\cite{blau}, the 
explicit symmetry breaking is expected to lead to a renormalizable one.

One can cast a perturbative 
framework  in BFYM, in order to check its $uv$-behaviour and in 
comparison with the standard perturbative expansion in YM. Actually there 
are two different ways to quantize BFYM theory and define Feynman rules. 
The first one is to regard the topological symmetry breaking term $g^2B^2$ 
belonging to the kinetic part of the lagrangian (\ref{uno.1}); in this way 
only the gauge symmetry needs gauge fixing and quantization and this is 
the case considered in this letter.
The second one regards the term  $g^2B^2$ as a true vertex; in this case the 
kinetic kernel is the same of the pure BF theory and requires gauge fixing 
and quantization also of the topological symmetry, although anomalous at 
the classical level. The procedure of quantization of the topological 
symmetry is quite involved, requiring a ghosts of ghosts structure due to
the reducible nature of the topological symmetry \cite{blau}. 
The anomalous term induces 
a dynamics also for the topological group degrees of freedom  which add
to the field content of the theory and compete with the topological ghosts 
to restore a local field theory; we will address to this case elsewhere.
\cite{fmtz}.

We then consider the ``minimal'' first order formulation and 
divide  the action (\ref{uno.1}) in a quadratic part and in 
a vertex one;
\ba  
L_0 &=&  i\varepsilon^{\mu\nu\alpha\beta}B^a_{\mu\nu}\pa_{\alpha}A^a_{\beta}
+B^a_{\mu\nu}B^{a\mu\nu}  \quad ,
\label{uno.6} \\
L_I &=& {i\over 2}g
f^{abc}\varepsilon^{\mu\nu\alpha\beta}B^a_{\mu\nu}A^b_{\alpha}
A^c_{\beta}\quad ,
\label{uno.7}
\ea 
where as usual the fields have been rescaled as 
$A\to gA$, $B\to B/g$  
in order to have the coupling constant on the vertex terms.

The BRS invariant action is obtained by adding to (\ref{uno.1}) the usual gauge 
fixing lagrangian, with the covariant gauge fixing condition
$\pa_{\mu}A_{\mu}=0$.
Feynman rules are read out of this lagrangian.

The kinetic terms display an off-diagonal structure and we obtain the 
following propagators in momentum space for the fields $A$ and $B$:
%
\vskip .4cm
\centerline{\vbox{\epsfxsize=12cm\epsfbox{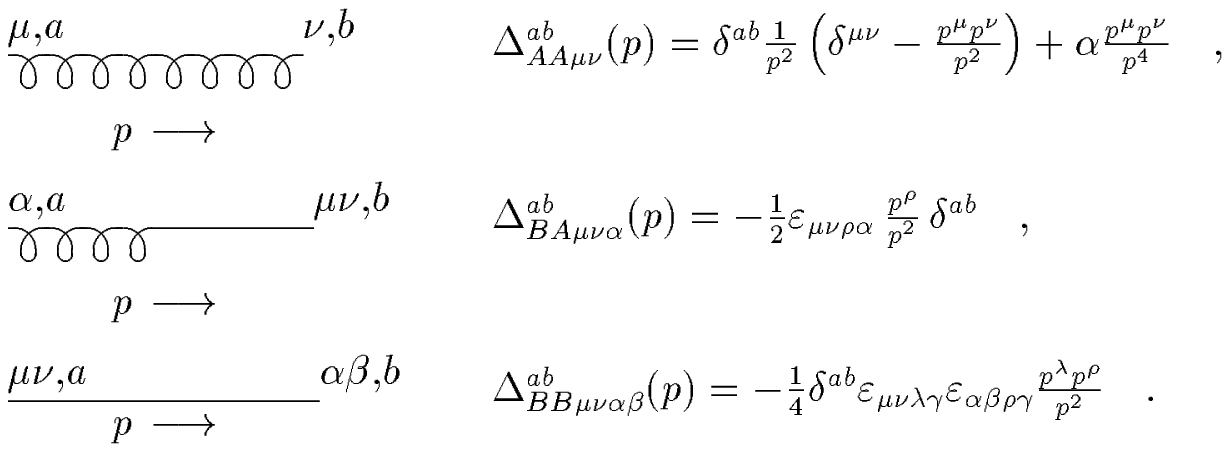}}}
\vskip .2cm
\centerline{\bf Fig. 1}
\vskip .3cm
%

The derivation of the propagators is sketched in the appendix ; in particular 
the transversal form of ${\Delta}_{BB}$, as displayed  in Fig.1, does 
not coincide with the one obtained by naive inversion of the kinetic operator. 
One must properly take into account the correction due to  
the spurious contribution of the topological zero modes which do not enter in 
the gaussian integration leading to YM theory; this point is discussed in 
the appendix, where the relative Ward identities are considered. 
Note the mass dimensions of the given propagators which accord with the
canonical  
scale dimensions of the fields; the propagator $\Delta_{BB}$ 
has dimension zero and 
behaves as a contact term at high momentum.

From (\ref{uno.7}) we see that BFYM has no self couplings for the gauge field 
$A$, the only relevant coupling being given by the vertex $BAA$: 
%
\vskip .4cm
\centerline{\vbox{\epsfxsize=10cm\epsfbox{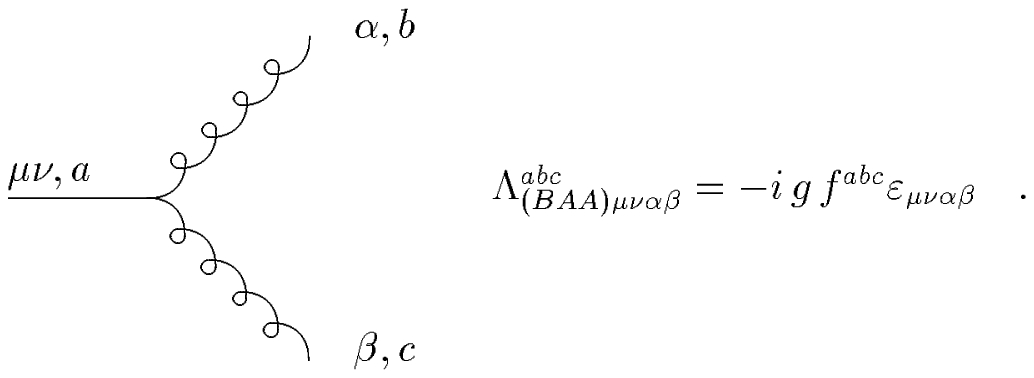}}}
\vskip .2cm
\centerline{\bf Fig. 2}
\vskip .3cm
%

As we will see is the off-diagonal structure of the propagators which 
reproduces the non linear self couplings of the gauge field.
To these Feynman rules we must add the usual ghost ones 
%
\vskip .4cm
\centerline{\vbox{\epsfxsize=9cm\epsfbox{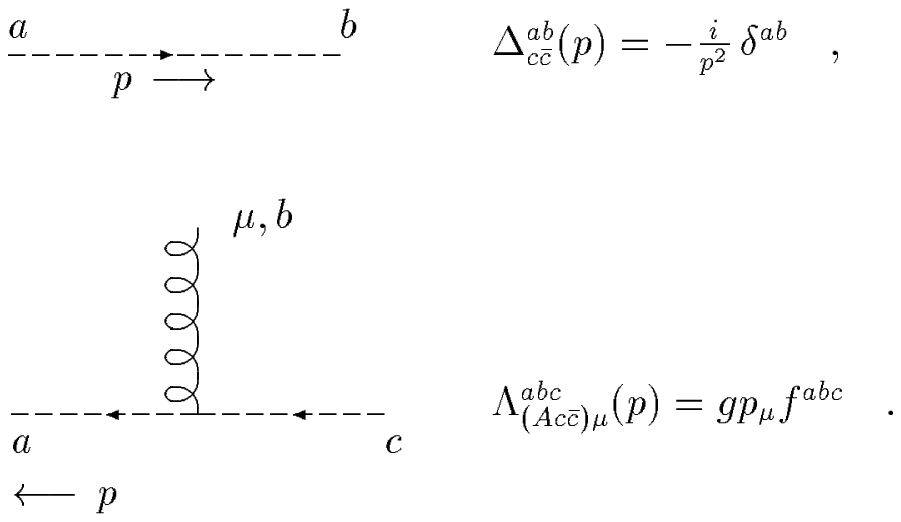}}}
\vskip .2cm
\centerline{\bf Fig. 3}
\vskip .3cm
%


\section{One Loop two point functions}
\setcounter{equation}{0}

In this section we consider the calculation of one-loop self-energies. 
In Fig. 4 are shown all the relevant one loop diagrams. The calculations 
are done in the Landau gauge ($\alpha =0$) using dimensional regularization, 
in dimension 
$D=4-2\epsilon$. In order to 
study the one loop renormalization and the $\beta$-function of BFYM theory 
we consider only the divergent parts of the diagrams considered.

The regularized divergent part of the self energies are then given by
\be 
\Pi_{AA}={1\over 6} g^2 z(\epsilon )\delta^{ab}
(p^2\delta^{\mu\nu}-p^\mu p^\nu ) 
\label{due.1}
\ee 
for the gluon self energy; by
\be 
\Pi_{AB}={3\over 4} g^2 z(\epsilon )\delta^{ab}
\varepsilon^{\alpha\beta\lambda\nu}p^\lambda
\label{due.2}
\ee 
for the $AB$ self energy; by
\be 
\Pi_{BB}=-{1\over 2} g^2 z(\epsilon )\delta^{ab}
I_{[\mu\nu ][\alpha\beta ]}
\label{due.3}
\ee 
for the $B$ self energy and by
\be 
\Pi_{c\bar c}={3\over 4} g^2 z(\epsilon )
\delta^{ab}p^2
\label{due.4}
\ee 
for the ghost self energy. 
$z(\epsilon )\equiv c_V{\Gamma (\epsilon)\over (4\pi)^2}$, where 
$c_V$ is the quadratic casimir for the 
Lie algebra of $SU(N)$.      
$I_{[\mu\nu ][\alpha\beta ]}=\delta^{\mu\alpha}\delta^{\nu\beta} 
-\delta^{\mu\beta}\delta^{\nu\alpha}$  is the antisymmetric identity.
%
\vskip .4cm
\centerline{\vbox{\epsfxsize=14cm\epsfbox{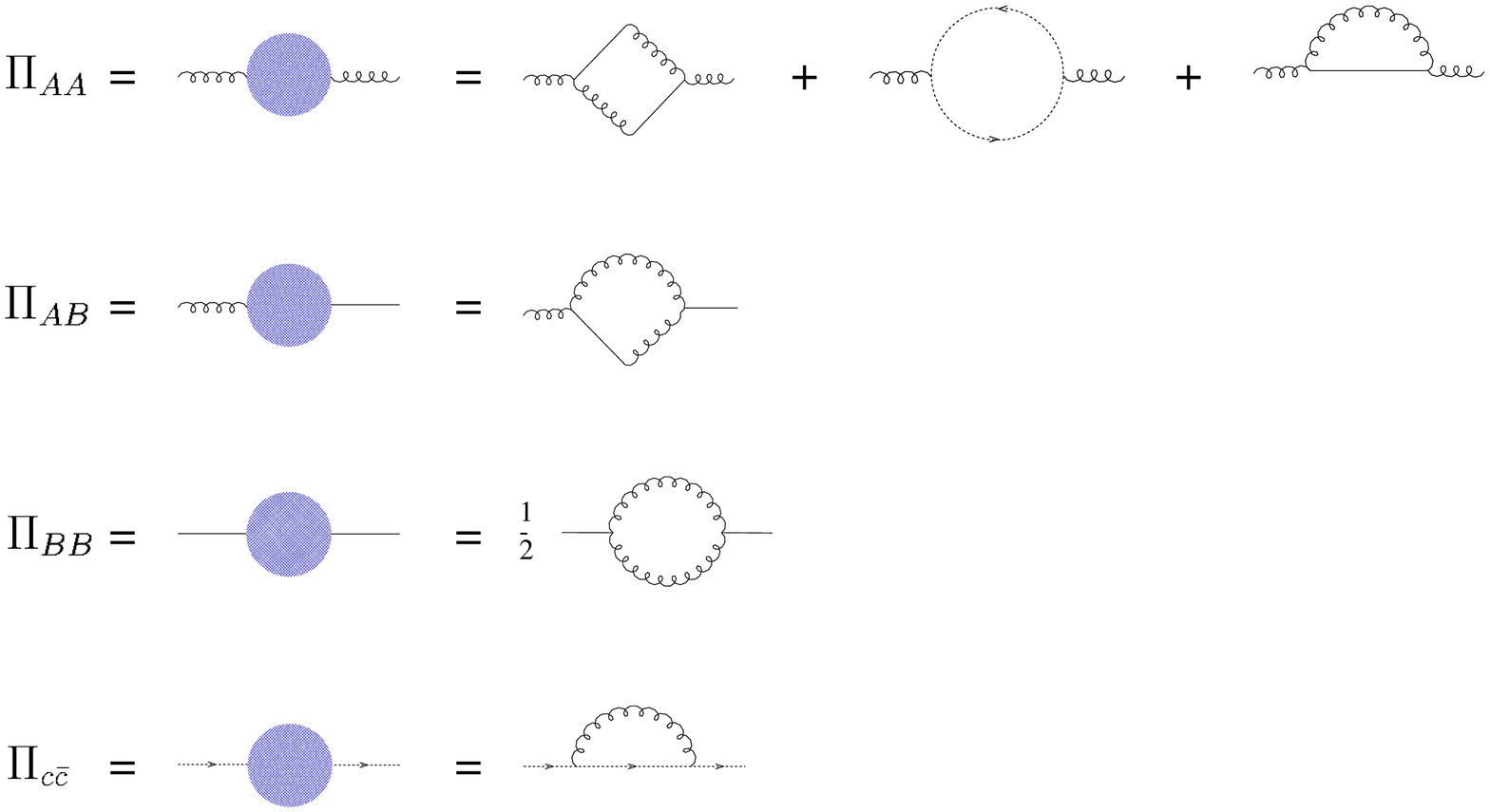}}}
\vskip .2cm
\centerline{\bf Fig. 4}
\vskip .3cm
%

As far as the self energies involving the $A$ and $B$ fields are concerned, 
note that they all contribute to the one loop two point functions. 
For example consider the correlator $<AA>$; in Fig. 5 we see how due to 
the structure of the propagator matrix  the 
one loop contribution to this function is recovered. 
%
\vskip .4cm
\centerline{\vbox{\epsfxsize=12cm\epsfbox{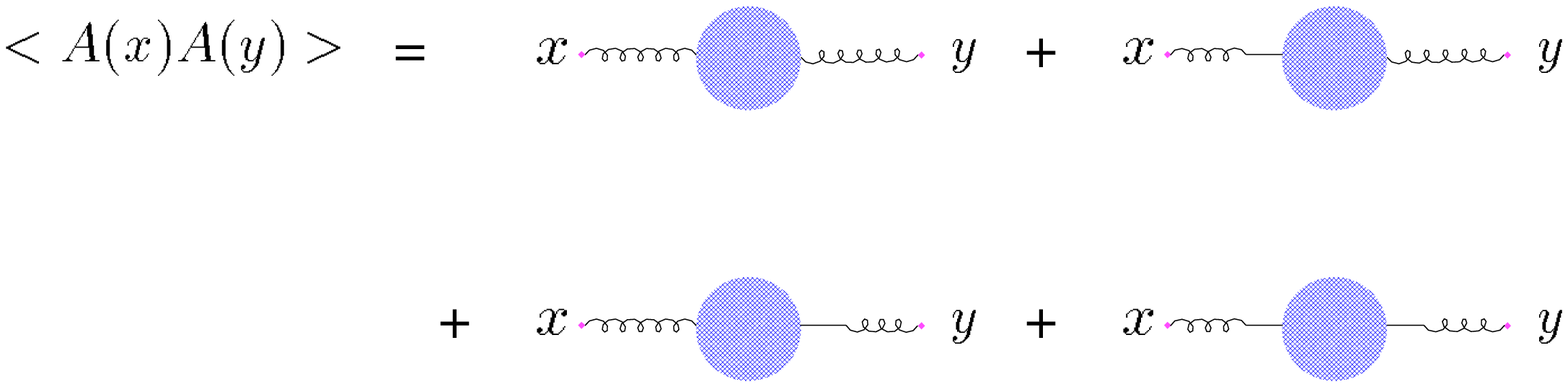}}}
\vskip .2cm
\centerline{\bf Fig. 5}
\vskip .3cm
%

In an analogous fashion it works for the two point functions $<AB>$ and 
$<BB>$.

\section{Vertex One Loop Diagrams}
\setcounter{equation}{0}

The superficial degree of divergence for the vertex diagrams is 
given by the formula 
\be 
\omega =4 - (E_A +  E_c) - 2E_B \quad ,
\ee 
where $E_A$, $E_B$ and $E_c$ represent the number of external legs 
joined to the diagram {\it via} $A$, $B$ and $c$ respectively. 
We then obtain for BFYM theory 
the four divergent vertex diagrams reported in Fig. 6.

The full calculation of these vertices should take into account more over 
sixty diagrams including permutations; 
we restrict the calculation  only to the divergent parts of the 
first two vertices. 
The divergent part of the ghost vertex $\Gamma_{Ac\bar c}$ is 
vanishing as in the standard calculations, owing to the 
transversality of the 
propagators in the Landau gauge. For the same reasons also the vertex 
$\Gamma_{BAA}$ is found to be finite at one loop order; this is the same 
vertex of the pure BF theory and seems to behave in a fashion corresponding to 
the topological theory. The last two vertices, $\Gamma_{AAA}$ and 
$\Gamma_{AAAA}$ 
do not belong to the tree level BFYM action and correspond to the 
nonlinear self interactions of YM which in this way are recovered into the 
theory. 
These vertices, joined to the gluon self 
energy (\ref{due.1}), originate from an $F^2$ term which the symmetries of the 
theory allow to enter in the quantum action; we will see in the next 
section how renormalization has to be performed in order to produce all the 
required counterterms.

%
\vskip .4cm
\centerline{\vbox{\epsfxsize=14cm\epsfbox{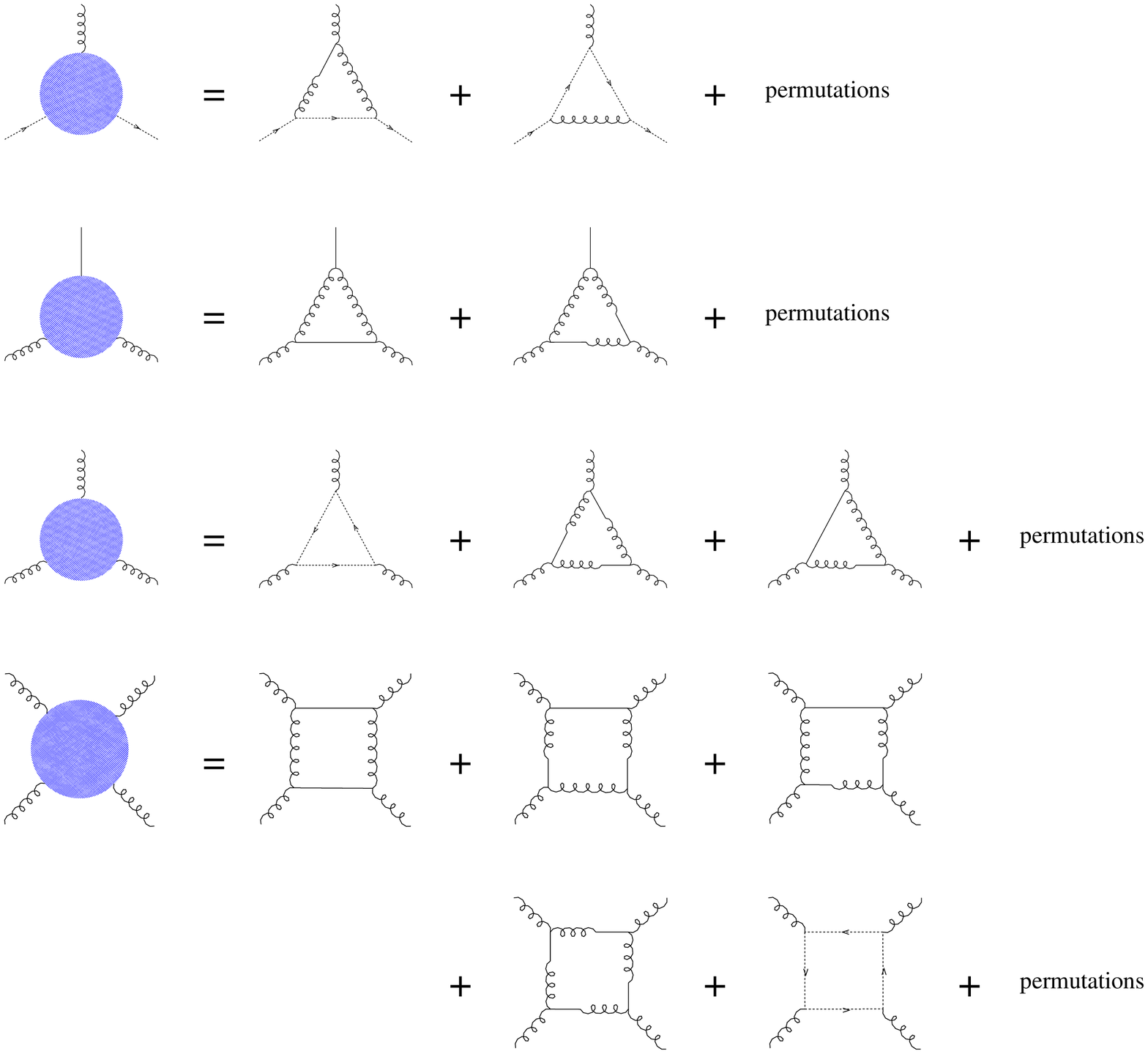}}}
\vskip .2cm
\centerline{\bf Fig. 6}
\vskip .3cm
%

\section{Renormalization and $\beta$-function}
\setcounter{equation}{0}

Renormalization is performed substituting the bare quantities with the 
renormalized ones,  where in general an operatorial mixing is allowed by 
the symmetry and parity properties of the fields. We then write
\ba  
B_0 &=& Z_{BB} B_R +iZ_{BA}*F_R\quad ,\nonumber \\
F_0 &=& Z_{AA} F_R\quad,
\label{qua.1}
\ea  
with $F_R=dA_R+Z_{AA}Z_g g_R [A_R,A_R]$, where $g_0= Z_g g_R$ and where 
the gauge Ward identities among renormalization constants have been imposed. 
Note that $B$ and $F$ have opposite parity; moreover a mixing of $B_R$ in 
$F_0$ is not allowed since $F$ must be a curvature tensor. 
This field mixing introduces the term $F^2$ absent at tree level in the 
theory and the counterterms relative to the gluon self-energy and 
to trilinear and quadrilinear gluon vertices.
We obtain 
\ba
S_{BFYM} &=& \int \tr [iB_0\wedge F_0 +B_0\wedge *B_0]\nonumber \\
&=& \int \tr [i Z_{BB}(Z_{AA}+2Z_{BA})B_R\wedge F_R +Z_{BB}^2B_R\wedge *B_R
\label{qua.2} \\ 
&& \qquad\qquad -Z_{BA}(Z_{BA}+Z_{AA})F_R\wedge *F_R ]\quad .\nonumber
\ea 
The renormalization of the ghost terms is performed in the usual way. 
Since Feynman rules of BFYM at tree level should not be modified we 
expect 
\ba
Z_{AA} &\simeq & 1+ ag^2_R z(\epsilon) +(...) 
+ o(g^2_R) \quad ,\nonumber \\
Z_{BB} &\simeq & 1+ bg^2_R z(\epsilon)+(...) 
+ o(g^2_R)\quad ,\label{qua.3}\\
Z_{BA} &\simeq &  cg^2_R z(\epsilon) +(...) 
+ o(g^2_R)\quad ,\nonumber
\ea 
where dots $(...)$ represent finite terms at order $g^2_R$. 
The value of the elements of the wave function 
renormalization matrix are assigned by 
direct comparison between the Feynman rules for the quadratic 
counterterms (\ref{qua.2}) and the divergent parts of the self-energies
(\ref{due.1}-\ref{due.4}). We obtain the following system
\ba 
Z_{BB}(Z_{AA}+2Z_{BA}) &=& 1+{3\over 4}g^2_R z(\epsilon)
+ (...) +o(g^2_R)\quad ,\nonumber \\
Z_{BB}^2 &=& 1-{1\over 2}g^2_R z(\epsilon)
+ (...) +o(g^2_R)
\quad ,\label{qua.4} \\
4(Z_{BA}^2+Z_{BA}Z_{AA}) &=& -{1\over 6}g^2_R z(\epsilon)
+ (...) +o(g^2_R)
\quad ;\nonumber 
\ea 
(note that the factor 4 in the third equation is due to the usual 
normalization for the $F^2$ term).
Solving the equations (\ref{qua.4}) at the order $g^2$ we find
\be
a={13\over 12}\quad ,\quad b=-{1\over 4}\quad ,\quad c=-{1\over 24}\quad .
\label{qua.5}
\ee 
The value of $a$ gives exactly the wave function renormalization for $A$ 
required in the Landau gauge for obtaining the correct 
value for the $\beta$-function of the theory. Indeed, introducing the 
ghost wave function renormalization, $c_0=Z_c c_R$,  
from (\ref{due.4}) we read 
\be 
Z_c=1+{3\over8}z(\epsilon )g^2_R + (...) +o(g^2_R)\quad ,\label{qua.5tris}
\ee 
and from the finiteness of the gluon-ghost vertex in 
Landau gauge we obtain 
\be
Z_gZ_{AA}Z_c^2=1+(...) +o(g^2_R) \quad ,\label{qua.5bis}
\ee  
where no divergent part at order $g^2_R$ is present. 
From (\ref{qua.5bis}) the renormalization of the coupling constant 
turns out to be  
\be
Z_g=1-{11\over 6}z(\epsilon )g^2_R + (...) +o(g^2_R)\quad ,
\ee
which gives $\beta_1=-{11\over 3}$ \cite{muta}. 
Therefore, as expected, the $uv$-behaviour of BFYM is the same of YM.
Also note that the values found in (\ref{qua.5}) 
give for the divergent part of the $BAA$ counterterm 
at $g^2_R$ level 
\be 
Z_{BB}(Z_{AA}+2Z_{BA})Z_{AA}Z_g= 1+(...) +o(g^2_R)\quad ,
\label{qua.6}
\ee 
according to the finiteness of $\Gamma_{BAA}$. After renormalization is 
performed is always possible to redefine $B_R$ in order to reabsorb the $F^2$ 
term and recover the tree level structure of the theory. Indeed defining 
\ba 
\tilde B_R &=& B_R +i\xi*F_R \quad ,\nonumber \\
\tilde F_R &=& F_R \quad ,\label{qua.7}
\ea 
where at $g^2_R$ order $\xi=Z_{BA}/Z_{BB}$,  
the renormalized action (\ref{qua.2}) becomes 
\be 
S=\int \tr [iZ_{BB}Z_{AA}\tilde B_R\wedge \tilde F_R 
+ Z_{BB}^2\tilde B_R\wedge *\tilde B_R]\quad . \label{qua.8}
\ee 
The transformation (\ref{qua.7}) gives a finite renormalization and, 
not involving the coupling $g_R$ contained in $F_R$, 
does not modify the correspondence with the  
renormalized Yang-Mills theory
\be 
S=\frac 14 \int\tr [Z_{AA}^2 F_R\wedge *F_R] \quad .
\ee 
In conclusion we have shown that this theory can be given a proper perturbative 
expansion and that the asymptotic free behaviour of BFYM 
coincides with that of YM. The perturbative formulation and the study of 
renormalization can be further investigated using algebric and cohomological 
tools and indeed the 3D BFYM theory 
has been studied in this way \cite{aamz} and the analysis will 
be extended to the 4D case. Some perturbative work on BF-type formulation of 
gravity theories can be found also in \cite{naka}. 

BFYM formulation opens the study to the relations between BF and gauge 
theories; in particular new non local observables can be introduced. These 
observables, describing topological higher linking numbers, where introduced 
in BF theories in \cite{horo} and can be naturally introduced in the gauge 
theory using the enlarged field content of BFYM. This investigation, discussed 
in \cite{fmz,mz} and previously started 
in \cite{ccgm,cm}, should be even more richer 
in the non minimal formulation \cite{fmtz} where the whole content of 
topological fields is present, added with new vectorial degrees of freedom, and 
is at most promising to produce a deeper understanding of the non perturbative 
sector of gauge theories.  
\section*{Aknowledgments}
M.Z. aknowledges  A. Grassi and M. Pernici for useful discussions and 
suggestions. 
The authors enjoyed useful discussions also with M.~Mintchev and 
are grateful to A. Accardi and A. Belli for checking some 
calculations. This work has been partially supported by MURST and by 
TMR programme  ERB-4061-PL-95-0789 in 
which M.~Z. is associated to Milan.
\appendix\section{Appendix} 
\setcounter{equation}{0}

In order to compute propagators consider $S_0 +\int J\Phi$, 
the quadratic part of the action (\ref{uno.1}) with the coupling of the 
fields to the external sources. Propagators are 
easily derived 
shifting for example the fields in momentum space by means of field 
independent functions, 
$A(p)$, $B(p)$, $b(p)$ $\rightarrow$ $A(p)+C_A(p)$, $B(p)+C_B(p)$, 
$b(p)+C_b(p)$, and solving for the 
$C$'s in such a way that linear terms in the fields disappear \cite{horne}. 
$b$ is the auxiliary field which implements the gauge fixing condition and 
playing a role only in the inversion of the kinetic operator. 
The corresponding solution is given by 
\ba
C_A(p)^a_\mu &=& -{1\over p^4}(p^2 J^a_{A_{\mu}}-p_\mu p^\nu J^a_{A_{\nu}}) 
-{1\over 2p^2}\varepsilon^{\alpha\beta\nu\mu}p_\nu J^a_{B_{\alpha\beta}} 
-i{p_\mu\over p^2}J^a_b  \quad ,    \nonumber\\
C_B(p)_{\mu\nu}^a &=& {1\over 2p^2}(p_\mu p^\alpha J^a_{B_{\nu\alpha}} +
p_\nu p^\alpha J^a_{B_{\alpha\mu}}) +{1\over 2p^2}
\varepsilon^{\mu\nu\alpha\beta}p_\alpha J^a_{A_{\beta}} \quad ,   \nonumber\\
C_b(p)^a&=& {i\over p^2}p^\alpha J^a_{A_{\alpha}} \quad ,      \nonumber
\ea 
and produces the propagators given in Fig. 1 with the exception of the 
$\Delta_{BB}$ term, which turns out to be
\be
\tilde\Delta_{BB\mu\nu\alpha\beta}^{ab}={\delta^{ab}\over 4p^2} 
(p^\nu p^\beta \delta^{\alpha\mu}+
p^\alpha p^\mu \delta^{\nu\beta}-
p^\nu p^\alpha \delta^{\mu\beta}-
p^\mu p^\beta \delta^{\alpha\nu}) \quad .\label{appendix.2}
\ee 
Note that this propagator is not transversal. Indeed one loop calculations of 
the self energy $\Delta_{BB}$ show that the correct structure is that reported 
in Fig. 1 and this fact agrees with what predicted by the Ward identity 
\be
\partial_\mu \partial_\alpha\Delta_{BB\mu\nu\alpha\beta}=0 \quad , 
\ee 
which can be derived by differentiation of the Ward identity on the 
connected Green functions generator functional.

To understand the mismatch between (\ref{appendix.2}) and $\Delta_{BB}$ 
note that in our 
treatment we have left undetermined the measure 
over $B$ using the naive one. Indeed the correspondence beween first and 
second order formalism should be written as 
\be
\int [\D B][\D A] e^{-S_{BFYM}} \simeq \int [\D A] e^{-S_{YM}}\quad ,
\ee 
where $[\D A]$ is the usual gauge fixed measure and $[\D B]$ is the measure 
over the orbits of the topological group (\ref{uno.5}). In our measure instead 
we have also the integration over the zero modes 
of the topological group, i.e. the configurations $B$ such that $B=D\eta$, 
$\eta$ 1-form, which are 
not coupled to $F$   
and do not contribute to the gaussian integration  
owing to the Bianchi identity. They 
give the overall factor 
\be 
\hat\Delta_{BB\mu\nu\alpha\beta}^{ab}={\delta^{ab}\over 4}
I_{[\mu\nu ][\alpha\beta ]}\quad .  
\ee 
This contact term is exactly the amount of the mismatch found, 
$\Delta=\tilde\Delta -\hat\Delta $. Therefore we have to take 
into account the presence of 
the spurious contribution of topological zero modes 
and assign to $\Delta_{BB}$ the correct tensorial 
structure following the Ward identity.

\end{document}